# Displacement Waves of Oxygen Atoms in Bi,Pb-2223 Lattice of Composites Annealed in Oxygen Reduced (O$_2$+N$_2$) Atmosphere

T.P. Krinitsina, S.V. Sudareva, E.I. Kuznetsova and Yu.V. Blinova

*Institute of Metal Physics, Ural Branch of Russian Academy of Sciences, Ekaterinburg, Russia*

Annealing of Bi,Pb-2223/Ag composites in (O$_2$+N$_2$) atmosphere at 820-780°C is believed to reduce the number of the accompanying phases, to make contacts between crystallites closer and to increase the critical current. The goals of this study were to reveal the changes in the 2223 lattice at annealing in the reduced oxygen atmosphere, to elucidate the reasons of these changes and to discuss their effect on the ceramics superconductivity. After the annealing the transversely-polarized displacement waves of oxygen atoms in [010]$_{2223}$ direction have been found in the 2223 phase by electron diffraction analysis. These waves could appear due to the lack of oxygen in the 2223 lattice or to the nitrogen penetration in it. As demonstrated by the X-ray photo-electron spectroscopy and nuclear microanalysis, nitrogen does not interact with the 2223 lattice, and the oxygen index decreases to 9.67, which is lower than the stoichiometric. Thus, the atomic displacement waves result from the lack of oxygen in Bi-O bilayers.

# I. INTRODUCTION

Annealing in the ($O_2$+$N_2$) atmosphere at 820-780°C is presently widely used in fabrication of Bi,Pb-2223/Ag composites. It is believed that at these annealing conditions the number of accompanying phases is reduced, contacts between crystallites improve and critical current increases. The goal of this study is to determine the character of changes in the 2223 phase lattice observed under the annealing in the reduced oxygen atmosphere, to elucidate the reasons of these changes and to discuss their effect on the ceramics superconductivity.

# II. EXPERIMENT AND DISCUSSION

The investigation has been carried out on 61-filament Bi,Pb-2223/Ag composites annealed at the last step of sintering in the mixture of (0.75%$O_2$ + 99.25%$N_2$) at the pressure of 10 bar. The methods of electron diffraction (JEM -200CX), X-ray electron microscopy (VG ES-CALAB MK II), nuclear microanalysis, X-ray photo-electron spectroscopy (XPES) and measurements of temperature dependences of magnetic moments (SQUID-magnetometer MPMS-XL-5) have been applied.

As demonstrated by transmission electron microscopy, in the majority of electron diffraction patterns of the 2223 phase with [001] zone axis the reflections of (0n0) type (where n is 1, 3, 5, etc) disappear in the [010]* row passing through the origin, whereas in other rows parallel to the [010]* direction but not passing through the (000) there is a complete set of reflections (Fig. 1). This is a characteristic feature of composites annealed in the reduced oxygen atmosphere. In case of composites heat treated in the air two complete perpendicular rows of reflections passing through the zero node are, as a rule, present in the electron diffraction patterns [1] (Fig. 2). The peer analysis of electron diffraction patterns shows that the absence of (010), (030), (050), etc. reflections in the [010]* row passing through the (000) node results from the transversely-polarized waves of displacements of oxygen atoms with the wavelength λ = $a$ = 54 nm directed along the **b** axis. Note that these transversely-polarized waves of O atoms arise in the same direction that the well known waves of displacements of Bi and Pd atoms [2]. It may be suggested that these displacements are inherent of oxygen atoms located in double Bi-O layers, that is in the places where the super-stoichiometric oxygen is usually arranged [3], the concentration of which is estimated from the value of δ ($Bi_2Sr_2Ca_{n-1}Cu_nO_{4+2n+\delta}$; n = 1, 2, 3).

Two models may be suggested to account for the formation of these waves of the oxygen atoms displacement. One of them attributes these waves to the lack of oxygen in the 2223 lattice when the oxygen index is lower than stoichiometry, that is lower than 10. In this case, according to [4], the atomic displacement waves arise to lower the electron energy of a system. The second model suggests that nitrogen atoms penetrate into the 2223 lattice and occupy the vacant oxygen positions in it, which is accompanied with changing of Cu valence and the formation of oxygen vacancies.

To chose which of these models is valid in our case, we should find out whether nitrogen inserts the 2223 lattice and what is the exact oxygen index in the 2223 phase after its annealing in the ($O_2+N_2$) atmosphere.

As demonstrated by XPES, within this technique sensitivity (less than 1 at. %) there is no nitrogen in the "volume" of the annealed HTSC Bi-ceramics (scraped off in the vacuum). Thus, it may be concluded that nitrogen does not interact with the 2223 lattice and it can't be the reason of the above described specific state of the lattice, and the latter is determined by the lack of oxygen. According to the data of nuclear microanalysis the oxygen index of the ceramics gets lower than stoichiometric (x = 10) and reaches the value of 9.67 ± 0.20.

An absence of any additional components in Bi4f spectrum obtained by XPES indicates that bismuth is present only in one chemical state. According to [5], oxygen is concentrated in Bi-O layers. Consequently, there is good reason to suppose that oxygen contents decreases down to (9.67 ± 0.20) at the expense of stoichiometric oxygen, and it is $Bi^{5+}$ that vanishes in the 2223 lattice whereas $Bi^{3+}$ is mainly retained. Besides, there is quite an intensive line with the bonding energy of 534.1-534.2 eV in the oxygen spectrum O1s which can be attributed to such bonds as C=O and O=C=O. According to [6], these bonds belong to the so-called functional groups (near-lattice formations) which retard oxygen diffusion and prevent the 2223 ceramics saturation with oxygen at annealing.

## III. CONCLUSION

The analysis of the results obtained shows that the transversely-polarized waves of displacements of oxygen atoms appear due to the lack of oxygen (unstable lattice). There are strong grounds to believe that oxygen content in the lattice decreases at the expense of stoichiometric oxygen located in double Bi-O layers, and it is in these layers that oxygen atoms are displaced due to their lack there.

## ACKNOWLEDGEMENTS

The specimens studied were fabricated by the method of "powder in tube" at Bochvar High-Technological Institute of Inorganic Materials. The structure and physical properties have been studied in the Collective Use Center at the Institute of Metal Physics Ural Branch of RAS and the X-ray photo-electron spectra were taken in the Collective Use Center at the Institute of Solid State Chemistry Ural Branch of RAS.

## REREFENCES

Figure Captions.

Fig. 1. Electron diffraction patterns of the ceramics of Bi,Pb-2223/Ag composite annealed in ($N_2 + O_2$) atmosphere; zone axis is [001], (010), (030)… reflections in the row $[020]^*$ are absent.

Fig. 2. Electron diffraction patterns of the ceramics of Bi,Pb-2223/Ag composite annealed in the air; zone axis is [001], both perpendicular rows of {100} type passing through the (000) node are seen.

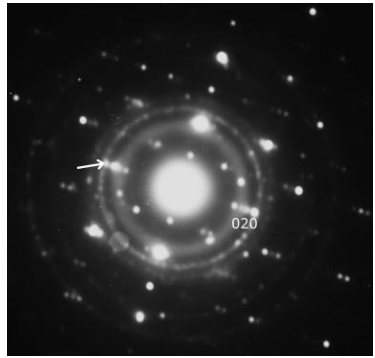

Fig. 1.

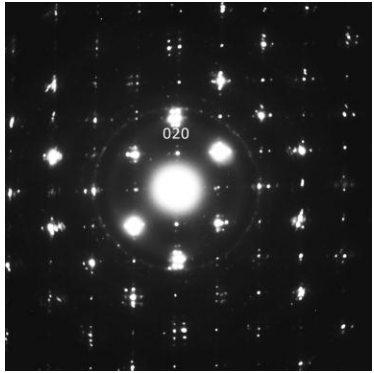

Fig. 2.